\documentclass[fleqn,usenatbib]{mnras}

\newcommand{\ramses}{\textsc{ramses}}
\newcommand{\agama}{\textsc{agama}}
\newcommand{\nexus}{\textsc{nexus}}

\usepackage{siunitx}
\usepackage{amssymb}
\usepackage{graphicx}

\title[Chaotic Orbits and Rapid Bulge Formation]{Turbulent gas-rich discs at high redshift: Chaotic orbits and rapid bulge formation at high gas fraction}
\author[Johan M. Villa-Alatorre et al.]{
Johan M. Villa-Alatorre,$^{1}$\thanks{E-mail: jvil0972@uni.sydney.edu.au}
Joss Bland-Hawthorn,$^{1}$
Thor Tepper-Garcia,$^{1}$
Oscar Agertz,$^{2}$ and \newauthor
Christoph Federrath$^{3}$
\\
$^{1}$Sydney Institute for Astronomy, School of Physics, A28, The University of Sydney, NSW 2006, Australia\\
$^{2}$Lund Observatory, Division of Astrophysics, Department of Physics, Lund University, Box 118, SE-221 00 Lund, Sweden\\
$^{3}$Research School of Astronomy and Astrophysics, Australian National University, Canberra, ACT 2611, Australia
}

\date{Accepted XXX. Received YYY; in original form ZZZ}

\pubyear{\the\year{2026}}

\begin{document}

\label{firstpage}
\pagerange{\pageref{firstpage}--\pageref{lastpage}}
\maketitle

\begin{abstract}

Using the \nexus\ N-body/hydrodynamic framework, we show that stellar bars that emerge in high-redshift disc galaxies are different in their orbital characteristics compared to their low-redshift counterparts. We identify bar-supporting orbits and, among them, find the fraction of chaotic orbits compared to regular orbits ($f_{\rm chaos}$). This fraction rises above 60\% for models with $f_{\rm gas} > 60 \%$; for $f_{\rm gas} \lesssim 60\%$, we observe chaotic fractions close to zero.
The emerging bars seen in the $f_{\rm gas}= 80\%$ and $100\%$ simulations collapse in $\lesssim$1 Gyr to form central bulges supported by rotation, and this collapse follows soon after the chaotic fraction starts to dominate ($f_{\rm chaos} \gtrsim 50\%)$. Moreover, the gravitational potential in those simulations steepens rapidly over that period, and by integrating the chaotic orbits, we find that orbits that deviate from the evolving mean field are more likely to be chaotic. Hence, we argue that the irregularity associated with the rapidly evolving, central potentials drives higher chaotic fractions in these gas-rich bars. As a result, the emergent bars collapse through `violent relaxation' to form oblate rotating bulges. Supporting evidence for this picture comes from the fact that particles of different mass share identical radial profiles, a natural consequence of the rapid relaxation process.
\end{abstract}

\begin{keywords}
galaxies: high-redshift, galaxies: formation, galaxies: evolution, galaxies: disc, galaxies: bar, galaxies: bulge
\end{keywords}

\section{Introduction} 
Modern astronomical observations at unprecedented redshifts like those from Atacama Large Millimiter/submillimeter Array (ALMA) and the James Webb Space Telescope (JWST) have been casting a new light on galactic disc formation in recent years. Remarkably, disc systems are detected to at least a redshift $z\approx 7.3$ \citep{row24} with a possible candidate at $z\approx 9.1$ \citep{tok22}.
Barred disc galaxies are observed at $z \gtrsim 4$ \citep{tsukui_detecting_2023,smail_hidden_2023,boogaard_stellar_2026} with one source recently discovered at $z\approx 5.1$ \citep{wang_massive_2026}. At the observed epoch, these systems were dominated by high gas fractions \citep{tacconi_phibss_2018}, very different from their low-redshift counterparts today. Little is known about the internal properties of high-redshift, gas-rich discs,  which needs to be addressed given that essentially all discs must have evolved through such a phase \citep{bland-hawthorn_turbulent_2024}.

In \cite{tepper-garcia_span_2024}, we introduced our parsec-scale simulation framework \nexus\ to study idealised galaxies evolving with feedback and turbulent gas over the full range of $f_{\rm gas}$ (see Eq. \ref{e:fgas}). 
In earlier papers, we reveal a rich variety of phenomena arising in turbulent gas-rich discs: three-armed spirals and radial shear flows associated with gaseous bars \citep{bland-hawthorn_turbulent_2024}, greatly enhanced stellar migration \citep{zhang_enhanced_2026}, the recoil of the baryonic potential with respect to the total potential \citep{bland-hawthorn_turbulent_2025}, the rapid formation of massive stellar bars \citep{bla26a}, and stronger solenoidal turbulence driving towards the shear-dominated central regions of galaxies \citep{WattEtAl2026}. In the same vein, we now introduce another phenomenon that has emerged from a study of the stellar orbit properties, i.e. rapid central bulge formation due to the onset of chaos.

The study of chaotic orbits has its roots in the classical dynamics of Newton, but major advances in the modern understanding of chaos started to appear after 1950. In particular, KAM theory, developed by Kolmogorov, Arnold, and Moser, established why some quasiperiodic orbits remain stable under perturbations while others become chaotic \citep[for a review, see][]{2002ocda.book.....C}. Orbital structure provides the dynamical skeleton of a galaxy. \citet{1964AJ.....69...73H} demonstrated how phase space can separate into regular and irregular, or chaotic, regions. Determining whether stars occupy regular or chaotic orbits can provide insight into how galaxies maintain their structure, redistribute angular momentum, and evolve. In particular, chaotic orbits can allow stars to explore regions of phase space that are inaccessible to regular orbits, potentially facilitating the redistribution of stellar populations and driving the evolution of a galaxy towards a more axisymmetric or spheroidal configuration \citep{merritt_chaos_1996}.

Two galaxies can look almost identical in starlight but have very different underlying orbital structures. Specifically, a galaxy's evolving stellar bar does not tell you automatically whether most of the stars are on regular bar-supporting orbits or whether a significant fraction occupy chaotic regions near resonances. Distinguishing these cases requires a direct analysis of the orbital dynamics.

For a Hamiltonian system, an orbit is classified as either quasiperiodic, i.e. confined to an invariant torus in phase space and characterised by a fixed set of frequencies, or chaotic, i.e.  nearby trajectories diverge exponentially and the orbit ergodically fills a volume rather than a surface. Several methods exist to make this distinction quantitatively; \cite{skokos_chaos_2016} provides an exhaustive review.

The most direct approach is the computation of Lyapunov exponents \citep{benettin_lyapunov_1980, wolf_determining_1985}, which measures the mean rate of exponential divergence between neighbouring trajectories. A positive maximum Lyapunov exponent is the defining signature of chaos, and Lyapunov analysis therefore serves as the natural validation benchmark for other methods. A complementary approach is frequency analysis \citep{laskar_chaotic_1990, laskar_measure_1992, laskar_frequency_1993}, which characterises each orbit by the set of frequencies it samples over time. Regular orbits produce a stable, discrete frequency spectrum, while chaotic orbits show frequency drift.

Of particular interest due to its simplicity and fast convergence, the General Alignment Index (GALI) method and its Smaller Alignment Index (SALI) variant to track the evolution of two or more initially distinct, normalised deviation vectors that follow the tangent dynamics of an orbit. Here we explore the rise of chaos using the SALI method because it can identify the moment an orbit becomes chaotic. Moreover, it has been extensively applied to galactic potentials including barred systems under different set-up conditions \citep{manos_studying_2007, patsis_chaoticity_2022, manos_orbit_2021, manos_interplay_2013, manos_regular_2011}.

The structure of the paper is as follows. In Sec.~\ref{s:style}, we describe the simulations in brief, the classification of bar supporter stars, and how the chaotic orbits are identified.
In Sec.~\ref{s:floats}, we present the results of our analysis, including the evolving morphology of the bars, as a function of $f_{\rm gas}$. We discuss our findings in Sec.~\ref{s:dis}, and conclude with remarks in Sec.~\ref{s:conclusions}.

\section{Methods} \label{s:style}
\subsection{Simulations}

We model isolated galaxies in dynamical equilibrium using the \ramses\ N-body/hydrodynamics code at benchmark resolution ($N \sim$ $10^7$ particles), with star formation and metal production enabled (Table~\ref{t:models}). The effective number of gas resolution elements is a factor of ten higher than the particle count. Each galaxy is represented by three components: a live dark matter halo, a massive pre-existing stellar disc, and a turbulent gas disc. The full \nexus\ framework additionally supports a hot coronal gas component that fills the halo and supplies the disc with smoothly accreting gas after cooling; that component is not activated here. The framework is described in detail in \cite{tepper-garcia_span_2024}.

Within \nexus, two parameters prove decisive for whether a massive stellar bar forms early. The first is the disc mass fraction,
\begin{equation}
\label{e:fdisc}
    f_{\mathrm{disc}} = \left( \frac{V_{c,\mathrm{disc}}(R_{\mathrm{e}})}{V_{c,\mathrm{tot}}(R_{\mathrm{e}})} \right)^2_{\!R_{\mathrm{e}}=2.2R_{\mathrm{disc}}},
\end{equation}
which measures the degree to which disc baryons dominate the gravitational potential over the dark matter halo at the effective radius $R_{\mathrm{e}} = 2.2\,R_{\mathrm{disc}}$, where $R_{\mathrm{disc}}$ is the exponential disc scale length and $V_c(R)$ is the circular velocity at radius $R$. The second is the gas mass fraction within the same aperture,
\begin{equation}
\label{e:fgas}
    f_{\mathrm{gas}} = \left( \frac{M_{\mathrm{disc,gas}}}{M_{\mathrm{disc}}} \right)_{\!R_{\mathrm{e}}=2.2R_{\mathrm{disc}}},
\end{equation}
where $M_{\mathrm{disc}}$ is the total disc mass and $M_{\mathrm{disc,gas}}$ is the gas contribution within that radius. Notice $f_{\mathrm{gas}}$ refers to the prescribed initial baryonic fraction in gas which naturally depletes over time as presented in \cite{bland-hawthorn_turbulent_2024}, their Fig. 3.

At high redshift, elevated gas fractions and surface densities drive vigorous star formation \citep{tacconi_evolution_2020}. The resulting stellar feedback---together with gravitational instabilities---sustains a highly turbulent interstellar medium, whose physics is described in \cite{agertz_toward_2013} and \cite{agertz_vintergatan_2021}. Our halo masses of order $10^{11}\,\mathrm{M}_\odot$ are characteristic of dynamical mass estimates from recent \textit{JWST} observations at $z \sim 1$--$6$ and are consistent with expectations for Milky Way progenitors at $z \approx 3$--$5$ \citep{bland-hawthorn_turbulent_2024}. For the massive progenitor models presented here, we adopt $R_{\rm vir} \approx 40\,\mathrm{kpc}$ and $\log\,(M_{\rm vir}/\mathrm{M}_\odot) \approx 11$, appropriate for this epoch.

Table~\ref{t:models} summarises our simulation suite. Holding the disc fraction fixed at $f_{\rm disc} = 50\%$, we vary the initial gas fraction across six values: $f_{\rm gas} = (0, 20, 40, 60, 80, 100)\%$. The \texttt{nac} designation in each filename indicates that halo accretion is suppressed in this subset; other \nexus\ runs with active accretion and varying $f_{\rm disc}$ exist, but their analysis is left for future work. Animations of each model's evolution are available on our project website.\footnote{\url{http://www.physics.usyd.edu.au/turbo_discs/}}

Models with $f_{\rm gas} < 100\%$ include a pre-existing stellar disc with old-star fraction $f_{\rm old} = 1 - f_{\rm gas}$. The $f_{\rm gas} \approx 100\%$ model retains a negligible seed stellar population ($f_{\rm old} \approx 1\%$). The parameter space explored is broadly consistent with CEERS-2112, currently among the best-characterised Milky Way progenitor analogues, observed at photometric redshift $z \approx 3.0$ \citep{cos23}.

\begin{table}
    \centering
    \caption{Overview of Galaxy Models. The DM host halo properties
    ($M_{\rm halo} = 10^{11}\,\mathrm{M}_\odot$, $R_{\rm vir} = 37\,\mathrm{kpc}$,
    and $r_s = 9.2\,\mathrm{kpc}$) are identical across models, and are consistent
    with a Milky Way progenitor analogue at $z \approx 3$. Columns: (1) fraction
    of mass in the baryonic disc (Eq.~\ref{e:fdisc}); (2) initial gas fraction
    (Eq.~\ref{e:fgas}); (3) simulation label; (4) initial stellar mass;
    (5) initial gas mass; (6) disc scale length.}
    \label{t:models}
    \begin{tabular}{lcccccc}
        \hline
        $f_{\rm disc}$ & $f_{\rm gas}$ & Label & $M_\star$ & $M_{\rm gas}$ & $R_{\rm disc}$ \\
        & & & ($10^8\,\mathrm{M}_\odot$) & ($10^8\,\mathrm{M}_\odot$) & (kpc) \\
        \hline
        0.5 & 0.0 & \texttt{fd50\_fg00\_nac}  & 112  & 0.0  & 1.8 \\
        0.5 & 0.2 & \texttt{fd50\_fg20\_nac}  & 88.8 & 22.2 & 1.8 \\
        0.5 & 0.4 & \texttt{fd50\_fg40\_nac}  & 66.6 & 44.4 & 1.8 \\
        0.5 & 0.6 & \texttt{fd50\_fg60\_nac}  & 44.4 & 66.6 & 1.8 \\
        0.5 & 0.8 & \texttt{fd50\_fg80\_nac}  & 22.2 & 88.8 & 1.8 \\
        0.5 & 1.0 & \texttt{fd50\_fg100\_nac} & 0.0  & 112  & 1.8 \\
        \hline
    \end{tabular}
\end{table}
\subsection{Sample selection}
\label{s:select}
For each simulation, we first determine an initial snapshot, at a time henceforth $t_{\rm ref}$, to select our samples and start classification runs. Throughout this work, we also distinguish the simulation time $t$ from the time elapsed since $t=t_{\rm ref}$ which we define as $\tau=t-t_{\rm ref}$. The choice of reference time is guided by the following requirements: 1) $t_{\rm ref} \geq \tau_{0.2}$, where the latter indicates the time when the bar strength crosses the threshold $\text{A}_2/\text{A}_0 = 0.2$ measuring the $m=2$ Fourier amplitude (not to be confused with $\tau$). The time from this definition of bar formation was shown to obey an inverse relation with $f_{\rm disc}$ in \cite{fujii_dynamics_2018}. The most constraining requirement: 2) there exists a reliable bar orientation $\varphi_{\text{bar}}$ measurement for all subsequent snapshots.

To calculate $A_2/A_0$, we employ the approach introduced by \cite{dehnen_measuring_2022} and the code therein, which we refer to as Dehnen's method. In short, the algorithm initially identifies the bar region from  overlapping annular bins, for each of which it conducts an azimuthal Fourier decomposition of the density distribution
\begin{equation}
\label{e:A2}
A_m =
\frac{\left|\sum_j \mu_j W(R_j)e^{-im\phi_j}\right|}
{\sum_j \mu_j W(R_j)},
\end{equation}

\begin{equation}
\psi_m =
-\frac{1}{m}\arg\left[
\sum_j \mu_j W(R_j)e^{-im\phi_j}
\right],
\end{equation}
where $\mu_j$, $R_j$ and $\phi_j$ are the mass, cylindrical radius and azimuth of particle $j$, and $W(R)$ is a radial window function. The code returns the normalised $m=2$ amplitude $A_2(R)$, the bar orientation $\psi_2$, pattern speed $\Omega_b$, etc, as independent measurements from a single snapshot. We pass the young stellar disc in this step and throughout this work, which consists of stars born from the self-consistent evolution of our simulations, except in the gas-free case where we use the previously existing disc. To extract $A_2$ from the binned $A_2(R)$, we average this array across the annulus identified by the code.
For $f_{\rm gas} \leq 60\%$ models, the time series of $\text{A}_2$ and $\Omega_b$, as well as the difficulties arising from applying Dehnen's method to gas-rich models are discussed in detail elsewhere \citep{bla26a}. For the most gas-rich pair ($f_{\rm gas} =80\%,100\%$), we discuss $\text{A}_2$ in Sec.~\ref{s:fourierFig} and present the time series from the companion paper.

Accurate bar orientations $\varphi_{\text{bar}}$ are necessary to identify which particles support the bar with the libration test described in Sec.~\ref{s:class}. By rotating the frame of the simulation to that determined by Dehnen method's $\psi_2$, we find the bar to occasionally jump from its fixed orientation starting from $f_{\rm gas} = 40\%$. Hence, starting from the estimated $\psi_2$, we correct to obtain $\varphi_{\text{bar}}$ by visual inspection, i.e. when the bar is stationary in the corrected frame. As found by others, visual corrections are the most reliable \citep{iles25}.

\begin{figure*}
\includegraphics[width=\textwidth]{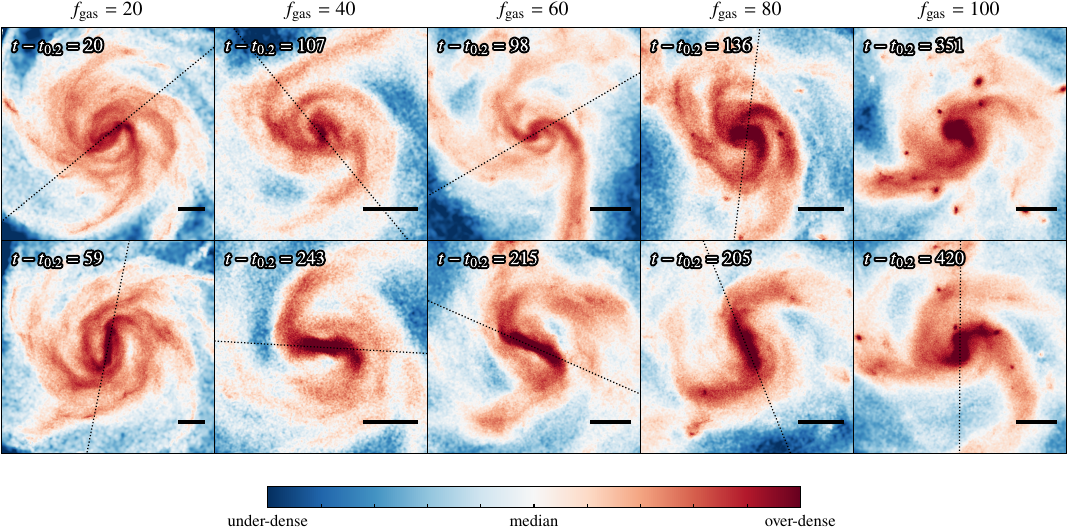}
\caption{Face-on density plot of the inner young stellar disc after the Fourier threshold $A_2/A_0> 0.2$ crossing for the simulations where conducting a libration test early is considered difficult. The time since threshold crossing is annotated for each panel in Myr, a dashed black line shows $\psi_2$ from Dehnen's method when it is returned, and a scale of 1 kpc is shown as a thick black line. Each panel is normalised to its median density. Top: At a time $t$ with $\tau_{0.2} < t < t_{\rm ref}$ that exemplifies the difficulty of determining a bar orientation. Bottom: At $t_{\rm ref}$, after which the orientation can always be well determined visually.}
\label{f:density}
\end{figure*}

Fig.~\ref{f:density} exemplifies some of the difficulties measuring the bar orientation (top) that drive our choice of reference time $t_{\rm ref}$ (bottom) beyond the typical formation threshold of $\tau_{0.2}$. In some of these cases (e.g.~for $f_{\rm gas} = 100\%$), while the bar first appears very early, it is later disrupted for a few snapshots as observed; thus delaying the initial snapshot selected. The resulting reference times are listed in Table~\ref{t:ref}. For $f_{\rm gas} \geq 60\%$, these were slightly adjusted to favour a common choice across simulations.

After this snapshot selection, we draw from it a sample of inner young disc stars. For this we pre-select all possible stars in the bar by introducing a radial cut, but this includes many circulating stars not trapped by the bar.
In some simulations, we find $R_{\rm bar}$ from Dehnen's method to oscillate between some adjacent snapshots even at the correct $\psi_2 = \varphi_{\rm bar}$; these $R_{\rm bar}$ measurements are presented in \citet{bla26a}. Moreover, even here, visual measurement is difficult due to the presence of flocculent spiral structure and  as illustrated in Fig. \ref{f:density}. 

Hence we opt to first run the libration test described in Sec. \ref{s:class} on a wide pre-selection to measure $R_{\rm bar}$ from stars co-rotating with the bar. We then take the final $R \leq 2 \, R_{\rm bar}$ star selection. Finally, from this population with between $4.0\times10^5$ and $1.1\times10^6$ young stars, we take a random sample of $10^5$ stars that we track and classify throughout this work in order to study if the emergence of chaos among them is related to structural changes in the bar. The characteristics of the selected snapshots are summarised in Table~\ref{t:ref}. 

\begin{table*}
    \centering
    \caption{Overview of selected snapshots. All particle counts in units of $10^6$;
    $N_{\rm star}$ is the total number of stars (young + pre-existing disc) at the
    reference snapshot, and $N_{\rm cut}$ is the number of young stars within
    $R_{\rm cut}$.}
    \label{t:ref}
    \begin{tabular}{ccccccc}
        \hline
        Label & $f_{\rm gas}$ & $\tau_{0.2}$ &
        \multicolumn{4}{c}{Reference snapshot} \\
        \cline{4-7}
        & [\%] & [Myr] &
        $t_{\rm ref}$ [Myr] &
        $R_{\rm cut}$ [kpc] &
        $N_{\rm star}$ &
        $N_{\rm cut}$ \\
        \hline
        \texttt{fd50\_fg00\_nac}  & 0   & 1295 & 1295 & 5.78 & 1.0 & 0.9 \\
        \texttt{fd50\_fg20\_nac}  & 20  & 586  & 645  & 3.40 & 1.5 & 0.4 \\
        \texttt{fd50\_fg40\_nac}  & 40  & 509  & 753  & 1.62 & 2.5 & 0.5 \\
        \texttt{fd50\_fg60\_nac}  & 60  & 254  & 469  & 2.66 & 3.0 & 1.1 \\
        \texttt{fd50\_fg80\_nac}  & 80  & 264  & 469  & 1.78 & 3.6 & 0.9 \\
        \texttt{fd50\_fg100\_nac} & 100 & 49   & 469  & 1.36 & 4.3 & 0.9 \\
        \hline
    \end{tabular}
\end{table*}

\subsection{Classification} 
\subsubsection{Bar supporters}
As mentioned above, we track bar orientations during our simulations to conduct a libration test. Simply, we convert each star's orbit from our selection to the rotating bar frame using the $\varphi_{\text{bar}}$ time series. Then, we identify the apocenters to determine whether their unwrapped angles in this frame, $\phi_{\text{apo}}$, remain locked to the bar: $\phi_{\text{apo, max}} - \phi_{\text{apo, min}} \leq \theta_h$(libration), or $\phi_{\text{apo, max}} - \phi_{\text{apo, min}} > \theta_h$(circulation), where $\theta_h$ is an arbitrary threshold we take as $\pi/3$. 

\begin{figure*}
\includegraphics[width=\textwidth]{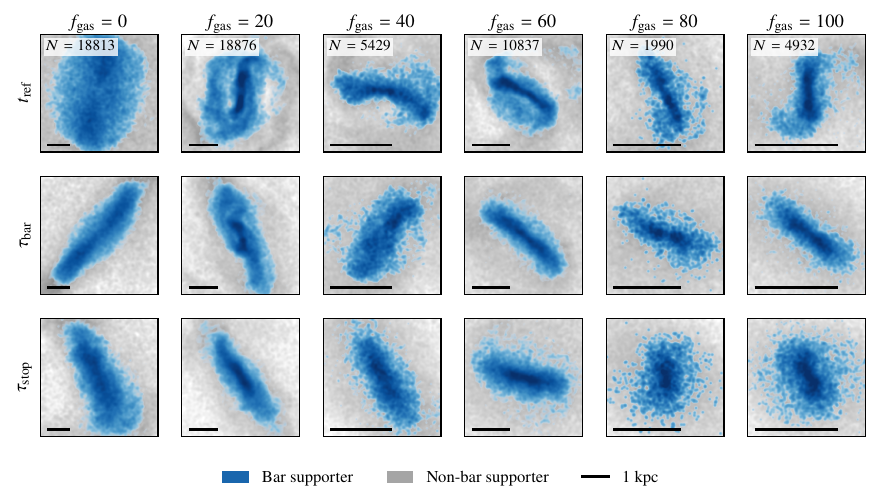}
\caption{Edge-on KDE map of bar membership classification at different epochs; with stars that passed the libration test (blue), and the rest of stars inside the $R < R_{\rm cut}$ selection region (gray). The number of bar supporters identified, $N_{\rm bar}$, is indicated as well as a scale of 1\,kpc. For every gas fraction, the field of view is set by the radius enclosing $98\%$ of bar supporters in the epoch when this radius is largest, and the same frame is used for all three rows. For any given panel, between $0.68\%-1.81\%$ of the blue population is outside the square view; hence they may be safely regarded as misclassifications. Top: The initial position of bar members in the reference snapshot of each simulation (at $t = t_{\rm ref}$).  Center: The same stars at the time $\tau_{\rm bar}$ when angle libration is evaluated. Bottom: Same stars at $\tau_{\rm stop}$, when the run to identify the chaotic population stops.}
\label{f:bar}
\end{figure*}

This wide margin suppresses error in our visual measurements and should be compensated by requiring a minimum of apocenters. Hence, we require stars to have at least 6 apocenters at the time the bar support is evaluated, which we refer to as $\tau_{\text{bar}}$, or otherwise they are discarded. This number was found to be the smallest so that $98\%$ of stars classified as bar supporters are within $R_{\rm bar}$ in all simulations as explained in the caption of Fig.~\ref{f:bar}. 

Moreover, our classification is permanent: i.e.~we do not directly track stars exiting the bar since visual measurements of $\varphi_{\rm bar}$ lose meaning as  bars start to collapse into bulges in the later snapshots of the most gas-rich simulations as discussed below. Hence, this classification is done only once in the middle of our runs at $\tau_{\text{bar}} = 500\,$Myr. As a consequence, we do not resolve bar supporters with period $ T_r > 83\,$Myr, which we discuss further in Sec.~\ref{s:dis}. 

The bar members classified by this method are shown at 3 different epochs in Fig.~\ref{f:bar}. First at $t_{\rm ref}$, then in the second row the same stars are plotted at $\tau_{\text{bar}}$, and the final row shows them at the end of the chaotic classification time $\tau_{\text{stop}}$. 
\label{s:class}

\subsubsection{Chaotic orbits}
\label{s:main}
To determine the fraction of chaotic orbits in our sample as a function of time we apply the Smaller Alignment Index (SALI) method \citep{skokos_alignment_2001} during the progression of each simulation. Briefly, this method allows us to identify chaotic motion by studying the alignment of deviation vectors $\hat{w_1}$, $\hat{w_2}$ along a reference orbit. This classification is possible since the SALI time series collapses to zero for chaotic orbits as explained below. Generally, if we use $|\cdot|$ to indicate the euclidean norm, the SALI value is defined at time $t$ as
\begin{equation}
\label{e:sali}
    \text{SALI}(t) = \text{min}\left(|\hat{w_1}(t) + \hat{w_2}(t)|, |\hat{w_1}(t) - \hat{w_2}(t)| \right) .
\end{equation}

When applied to N-body simulations, the timescale has been chosen in the order of 10\,Gyr of integration time within a series of individual frozen snapshots \citep[e.g.][]{chaves-velasquez_boxy_2017}. This autonomous Hamiltonian approximation tries to identify which orbits would be chaotic if one could ignore the evolving potential and thus ``extrapolate'' orbits indefinitely into the future by integration. Another implementation is the multi particle method of \cite{many_manda_efficient_2025} which does not evolve the deviation vector's variational equations, but uses nearby orbits; however, achieving low error requires very strict energy conservation.

The SALI (or GALI$_2$) method has also been applied to evolving barred Hamiltonians \citep[cf.][]{ manos_interplay_2013, manos_orbit_2021}, and chaos from time-dependent analytical potentials fitted to N-body simulations has been studied as well \citep[e.g.][]{manos_chaos_2014}. However, no approach in the literature operates on live simulations to our knowledge: as noted in \cite{manos_interplay_2013} the frozen-snapshot approach is favoured due to the live orbits alone missing information to evolve deviation vectors. In contrast, here we implement SALI to track the N-body trajectories of our selected particles  by ``interpolating'' the orbits between snapshots using integration to obtain the missing information. This is done in order to study dynamics subject to secular evolution. In practice, this means we are interested in identifying orbits that onset to chaos only on timescales relevant for bar evolution ($\approx 1\,$Gyr), rather than in the order of the Hubble time ($\approx 10\,$Gyr). The latter would normally be classified as `sticky orbits.'

To accomplish this, we first initialise two random, orthonormal deviation vectors per star from our selected sample, and apply the following procedure until a time $\tau_{\text{stop}} = 1\,$Gyr: Let $\vec{x}(t)$ be the inertial frame N-body state of these stars. For every adjacent pair of snapshots, we integrate the current $\vec{x}(t)$ during a time corresponding to that between snapshots. The snapshot cadence is typically $\Delta t \approx 10\,$Myr except for the gas-free model with $\Delta t \approx 20\,$Myr.

Integration was performed with the galactic dynamics library \agama\  \cite{vasiliev_agama_2019}, which directly supports computation of deviation vectors through the variational equations by integrating orbits with the parameter \texttt{der$\;=\;$True}. This process returns the transition matrix $\mathcal{M}$ between states from initial conditions and a potential. Here, the potential is constructed for each snapshot from the contribution of each individual component (e.g.~ gas, dark matter, young disc, etc), whose respective Fourier or multipole coefficients are linearly interpolated between snapshots to obtain a time-dependent potential. Finally, the new states returned by \agama\ are discarded in favour of the next snapshot state. Therefore, the purpose of integration is exclusively to produce the map $\mathcal{M}$ between the current deviation vectors $\hat{w}_i(t)$ to the next snapshot: $\vec{w}_i(t + \Delta t) = \mathcal{M} \, \hat{w}_i(t)$, and then the SALI is computed from Eq. \ref{e:sali} after normalising. Before discarding the integrated states from \agama, we compute the relative error between this result and the next snapshot state, normalising by the advanced arc of each orbit. Thus, we compare the chaotic fraction in each simulation binned by the average integration error experienced per star throughout the run as a diagnostic of whether our implementation is inducing chaos when forcing to track the simulation.

Having the SALI time series for each star, the final step is classification. For a detailed presentation of the method, the reader is referred to \cite{skokos_smaller_2016}. The evolution of this quantity can be shown to be \citep{skokos_detecting_2004}:
\begin{equation}
\label{e:sali2}
    \text{SALI}(t) \propto e ^ {-\left(\lambda_1 - \lambda_2 \right)\, t },
\end{equation}
where $\lambda_1\geq \lambda_2$ indicate the largest Lyapunov exponents. Thus, this scalar exhibits two distinct behaviours: for chaotic orbits, it decays exponentially towards zero on a characteristic timescale $\tau_{\text{SALI}} \approx (\lambda_1 - \lambda_2)^{-1}$; for regular orbits, or in cases where the two largest Lyapunov exponents coincide, it remains larger than zero and exhibits oscillatory behaviour. Since we initialise orthonormal deviation vectors, from Eq. \ref{e:sali} it can be seen that $\mathrm{SALI}(0)=\sqrt{2}$, thus every particle starts classified as a regular orbit with the same SALI, regardless of its history. 

Given our integration time is shorter than studies aiming at capturing sticky orbits, the SALI does not have sufficient time to decay to machine precision ($\approx 10^
{-16}$) or commonly adopted thresholds such as $10^{-8}$. We therefore adopt a wider threshold and instead demand SALI to cross it permanently rather than instantaneously for an orbit to be classified as chaotic. In order to determine this threshold we perform a test run of our code for $1$ Gyr in axisymmetric and spherical potentials, finding that SALI $\leq 10^{-3}$ misidentifies at most $1.4\%$ of orbits as chaotic. 

From the threshold crossing time we find the fraction of orbits that have been established chaotic as a function of time, which we name $f_{\text{chaos}}$ throughout this work, and it will monotonically increase by the permanent crossing definition.

\subsection{Origin of chaos and bar collapse}
We investigate some mechanisms behind some gas-rich bars weakening, a phenomenon we study in this work. First, we show this weakening through the evolution of the axis ratios of the bar, and use them to characterise the bulge-like structure that develops. The in-plane ($b/a$) and vertical ($c/a$) axis ratios are computed over time from our previously classified bar supporters. Here, we trim the outer $2\%$ likely misclassifications and compute the axis ratios using the halo ellipticity approach of \cite{dubinski_structure_1991}. This approach consists of computing the inertia tensor weighted by radius,
\begin{equation}
    I_{ij} = \sum_n \frac{x_{i,n} x_{j,n}}{r^2_n},
\end{equation}
with $r_n = a \sqrt{x^2_n/a^2 + y^2_n/b^2 + z^2_n/c^2}$ as the ellipsoidal distance. The method consists of calculating $I_{ij}$ from an initial guess of the axis ratios until their convergence is achieved up to a small tolerance \citep[for an application to N-body simulations see][]{allgood_shape_2006}. To measure $R_{\rm bar}$, we calculate the instantaneous radius of the $98^{\text{th}}$ percentile bar member in each snapshot and verify with a visual measurement once.

We follow the evolution of the total potential $\Phi$ during the classification runs to investigate its correlation with bars developing an axisymmetric morphology. Moreover, we compare the chaotic fraction to a measure of bar strength. \cite{combes_formation_1981} introduced the profile
\begin{equation}
    \label{e:Q_T}
    Q_{\mathrm{T}}(R) = \frac{F_{\mathrm{T}}^{\max}(R)}{\langle F_{R}(R) \rangle}
    = \frac{\max_{\varphi} \left| \partial \Phi / \partial \varphi \right|_{R}}
           {R \, \left\langle \partial \Phi / \partial R \right\rangle_{\varphi}} ,
\end{equation}
where $\langle \cdot \rangle_{\varphi}$ denotes an
azimuthal average, and both forces come from the \agama\ 
potential. The azimuthal average is equivalent to the $m = 0$ Fourier coefficient of $F_{R}$, so the denominator is the axisymmetric component of the radial force.
However, we find that this $Q_{\mathrm{T}}(R)$ oscillates significantly between snapshots. This is largely improved by symmetrising the force field by averaging over 180 degrees in azimuth. It can be shown the following is equivalent to removing the odd Fourier modes from the force term
\begin{equation}
F_{\mathrm{T, sym}}^{\max}(R)
=
\frac{1}{R}
\max_{\varphi}
\left|
\frac{1}{2}
\left[
\frac{\partial \Phi(R,\varphi)}{\partial \varphi}
+
\frac{\partial \Phi(R,\varphi+\pi)}{\partial \varphi}
\right]
\right|,
\end{equation}
which we use in place of $F_{\mathrm{T}}^{\mathrm{max}}(R)$ in Eq. \ref{e:Q_T}. Finally, the maximum of $Q_{\mathrm{T}}(R)$ within $R_{\rm bar}$, $Q_{\mathrm{b}}$ as introduced by \cite{buta_dustpenetrated_2001} is a common measure of bar strength we track over the simulations (notice that $F_{\mathrm{T, sym}}^{\max}(R) \leq F_{\mathrm{T}}^{\mathrm{max}}(R)$ and thus our $Q_{\mathrm{b}}$ are lower bounds on the standard definition):
\begin{equation}
    \label{e:Q_b}
    Q_{\mathrm{b}} = \max \left[ Q_{\mathrm{T}}(R) \right]_{ R < R_{\rm bar}}.
\end{equation}

\section{Results} \label{s:floats}
\subsection{Bar evolution}
We now present the results that are directly related to the observed change of morphology of the bars from different perspectives. 
\subsubsection{Morphological evolution}
\begin{figure*}
\includegraphics[width=\textwidth]{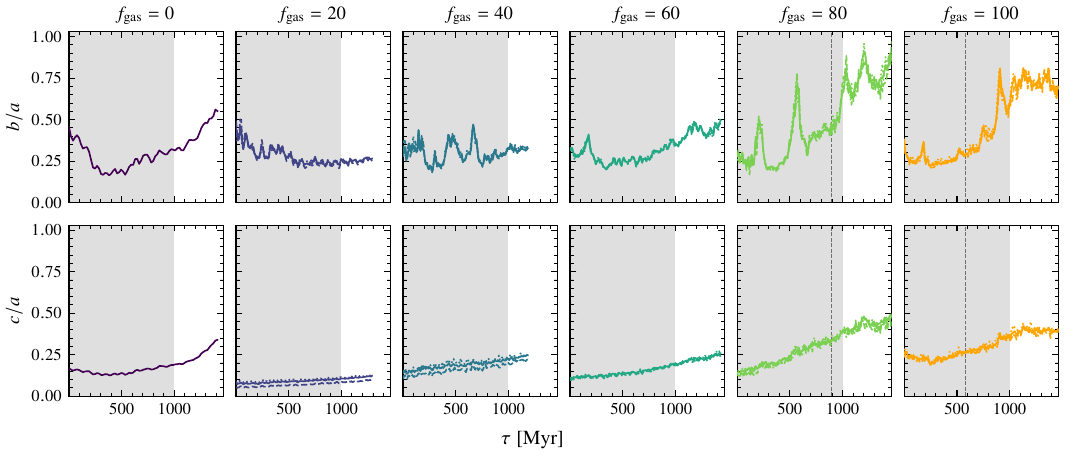}
\caption{Axis ratios from the classified bar supporter population of stars over time. Top: The in-plane minor to major axis ratio $b/a$. Bottom: The vertical ratio $c/a$. The time axis $\tau=t-t_{\rm ref}$ where $t$ is the simulation time and $t_{\rm ref}$ is defined in Sec. \ref{s:select}.
The time span for which the evolution of the chaotic fraction is estimated ($\tau \in [0,1000]$ Myr) is indicated as a gray shaded area. A vertical, thin dashed line indicates the time when the chaotic fraction $f_{\rm chaos}$ rises above $50\%$ of bar supporters for the models that surpass this share of chaotic orbits. Additionally, for models with star formation line styles  indicate particle selections according to their mass at $\tau = 0$: all particles (solid), top 85 per cent by mass (dashed), and bottom 15 per cent by mass (dotted).}
\label{f:axis}

\end{figure*}

First, we show the morphological analysis of the bar from the stars that were classified as librating. The last two panels of Fig.~\ref{f:bar} show the stars conforming the bars in the $80\%$ and $100\%$ gas simulations have lost their elongated shape by $\tau = 1000$ Myr ($t = 1466$ Myr). We show the evolution of the axis ratios $b/a$ and $c/a$ from these classified particles in Fig.~\ref{f:axis}.

The $b/a$ ratio defines the in-plane circularity of the disc, such that $b/a < 1$ indicates the disc retains some in-plane ovoid distortion. The $c/a$ ratio describes the degree of flatness seen when the disc is edge-on, such that the ellipticity of the edge-on projection is defined as $\varepsilon=1-c/a$.

As observed, the $f_{\rm gas} = 80\%,100\%$ bars eventually become close to axisymmetric in the plane while they also thicken vertically at a slower rate. Moreover, the time when the share of chaotic bar orbits crosses $50\%$ (presented in Sec.~\ref{s:chaos}) precedes the maximum of these curves. The $f_{\rm gas}=0\%,60\%$ models also show an upward $b/a$ trend, however, they never lose their bar shape. We also find little difference in the evolution of axis ratios when computing them using different selections according to particle mass. This is relevant for violent relaxation as discussed in Sec. \ref{s:relaxation}.

\begin{figure}
\includegraphics{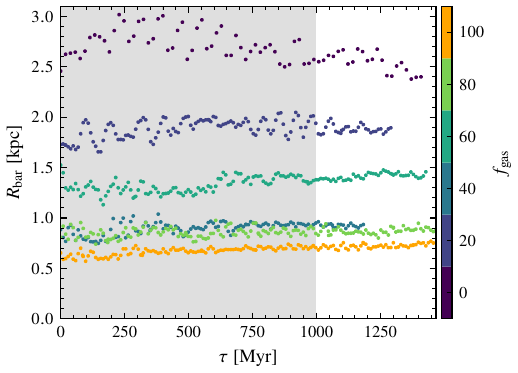}
\caption{The radial extent  $R_{\rm bar}$ from the $98^{\rm th}$ percentile of radii of the stars classified as bar supporters as a function of time. A shaded area indicates the window where the chaotic fraction was computed. Each curve is coloured according to its gas fraction. The time axis is defined in Fig.~\ref{f:axis}.}
\label{f:Rbar}

\end{figure}

The evolution of the radial extent of the bar is presented in Fig.~\ref{f:Rbar}. The general trend is that models richer in gas form progressively smaller bars as discussed in \citet{bla26a}, with the exception of the $f_{\rm gas} = 40\%$ bar which is smaller than expected. We find no significant evolution of $R_{\rm bar}$, even as the gas-rich bars weaken. 

\subsubsection{Fourier density}
\label{s:fourierFig}
\begin{figure}
\includegraphics{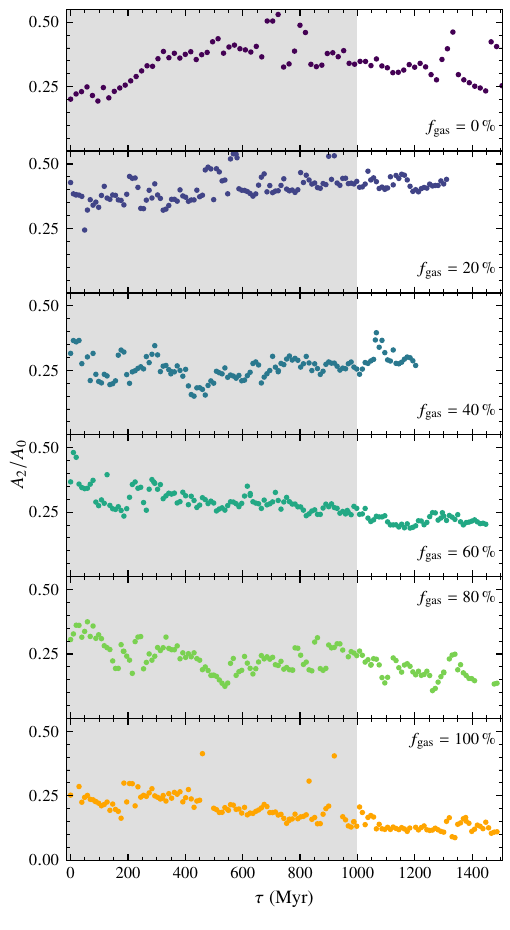}
\caption{The evolution of the normalised $m = 2$ Fourier density mode produced from Eq. \ref{e:A2}. The curves corresponding to $f_{\rm gas} \leq 60 \%$ are discussed in \citet{bla26a}. A shaded area indicates the window where the chaotic fraction was computed. Each curve is coloured according to its gas fraction, which is annotated in each panel. The time axis is defined in Fig.~\ref{f:axis}.
}
\label{f:A2}

\end{figure}

The normalised Fourier amplitude $A_2$ is shown in Fig.~\ref{f:A2}. As observed, the $0\%$ and $60-100\%$ curves show a downward trend during the classification runs, suggesting weakening bars. This is consistent with the same curves showing a rising $b/a$ (less flattening) in Fig.~\ref{f:axis}. As expected in bar evolution, the $A_2$ modes grow to a peak amplitude after an exponential increase, after which they saturate before declining. In fact, the gas-dominant simulations start to decay even before the chaotic classification window ($\tau < 0$).

\subsubsection{Gravitational potential}
Analysis of the gravitational potentials is an approach to show a possible connection between the secular evolution of the galaxies (i.e changes in the potential) and the observed change of bar morphology (relaxation) by means of increasing the chaotic fraction. Fig.~\ref{f:potentials} shows the evolution of the total potentials throughout the run. In this interval, the $f_{\rm gas}$ between $40\%-100\%$ potentials steepen rapidly, followed by loss of ellipticity in the $80\%, 100\%$ runs, and the collapse appears to occur for this pair between $\tau > 820$ Myr and $\tau < 1000$ Myr which is consistent with the peak in $b/a$ (Fig.~\ref{f:axis}) for these bars. By the end of chaotic classification, the $f_{\rm gas} = 60\%$ potential appears to have also gained some degree of axisymmetry, however, its bar never dissolves. The gas sub-dominant models do not exhibit more axisymmetry over time.

\subsection{Chaotic fraction}
\label{s:chaos}
\begin{figure*}
\includegraphics[width=\textwidth]{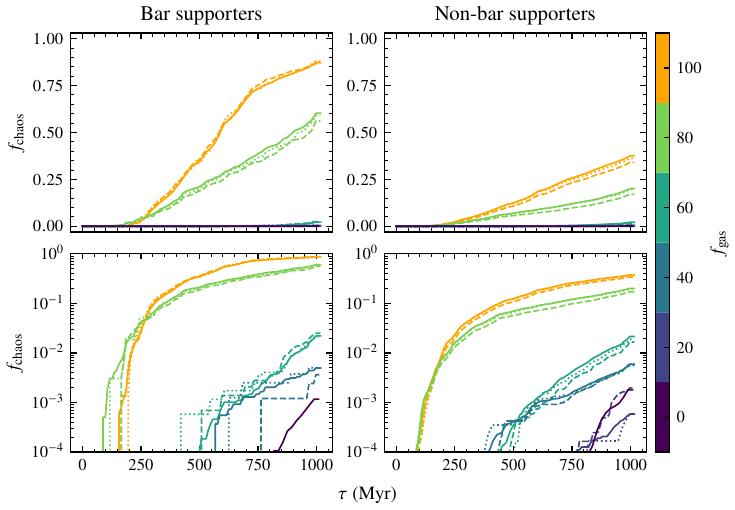}
\caption{Fraction $f_{\rm chaos}$  of orbits that permanently crossed the chaotic classification threshold  as a function of the time $\tau$ since each simulation's reference time for different groups of particles. Left: For stars that passed the libration test. Right: For stars that failed the libration test. Line styles further indicate particle selections according to their mass at $\tau = 0$: all particles (solid), top 85 per cent by mass (dashed), and bottom 15 per cent by mass (dotted). The top and bottom rows show the same plot in linear and log scale, and the curves are coloured according to the initial gas fraction of the simulation.}
\label{f:result}
\end{figure*}

We present the progression of chaotic onset for the populations according to their bar membership category and mass in Fig.~\ref{f:result}. Left panels (bar members) show the most gas-rich bars develop widespread chaos. In the $f_{\rm gas} = 100\%$ simulation, the chaotic population grows to $\approx 85 \%$ of bar supporters. For the $f_{\rm gas} = 80\%$ bar,  $f_{\rm chaos}$ climbs to $\approx 60 \%$ by the end of the run; and these populations cross $50 \%$ chaos after $590$ and $900$ Myr since the reference snapshot, respectively. As best observed in the bottom panels, the curves cluster in two different regimes separated about  $f_{\rm gas} \approx 70 \%$, and there is a general trend that increased gas presence yields higher chaotic fractions. 
These general trends are also observed in the right panels (non-bar members). Furthermore, the bar population in $80,100\%$ gas models is seen leading in chaotic fraction over the rest of the inner young stellar disc. This suggests the onset of chaos is not random in the most gas-rich models, it preferentially favors bar members (or stars deep in the potential) by up to $40\%-50\%$ until the orange curve in the left panel reaches a saturation stage after $700$ Myr, and from there it grows at the same rate as in the right panel.

We show the best between linear or exponential fits to the overall chaotic fraction from all stars in Table \ref{t:fits}. The $f_{\rm gas} \leq 60 \%$ chaotic fractions match an exponential curve whereas the most gas-rich pair fit best to a linear model. However, if we limit the fit to the first $200$ Myr, the $80\%$ and $100\%$ gas curves also match an initial exponential growth phase with ${\cal R}^2 > 0.98$; then, the process transitions to a linear increase, and a saturation phase at $\tau \gtrsim 700$ Myr only in the $100\%$ gas curve. The curves in Fig. \ref{f:result} also distinguish contributions from low and high mass particles. As presented, we find no significant difference in chaotic fraction among low v.s high mass particles. This is further discussed in terms of violent relaxation in Sec. \ref{s:relaxation}.

\begin{table}
    \centering
    \caption{Linear ($y = ax + b$) or exponential ($y = a\,e^{bx}$) fits to the
    full $f_{\rm chaos}$ time series for all sampled stars. Only the model with
    the highest coefficient of determination $\mathcal{R}^2$ is listed in each
    case.}
    \label{t:fits}
    \begin{tabular}{ccccc}
        \hline
        $f_{\rm gas}$ & Form & $a$ & $b$ & $\mathcal{R}^2$ \\
        \hline
        0   & exp & $8.3\times10^{-9}$  & $1.2\times10^{-2}$ & 0.989 \\
        20  & exp & $2.0\times10^{-7}$  & $8.4\times10^{-3}$ & 0.996 \\
        40  & exp & $3.5\times10^{-5}$  & $5.1\times10^{-3}$ & 0.994 \\
        60  & exp & $1.8\times10^{-5}$  & $6.9\times10^{-3}$ & 0.997 \\
        80  & lin & $2.0\times10^{-4}$  & $-2.7\times10^{-2}$ & 0.976 \\
        100 & lin & $4.3\times10^{-4}$  & $-6.4\times10^{-2}$ & 0.974 \\
        \hline
    \end{tabular}
\end{table}

\begin{figure}
\includegraphics{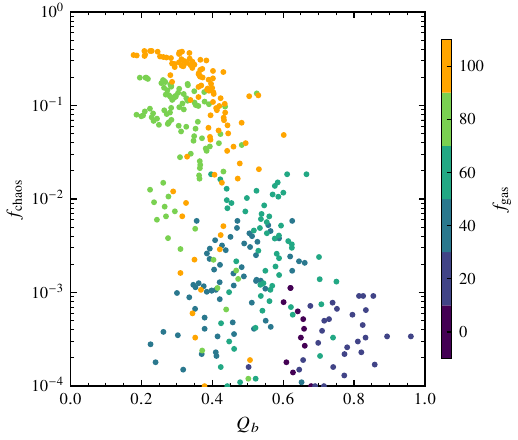}
\caption{Fraction of chaotic orbits as a function of the bar strength $Q_{\mathrm{b}}$ (Eq. \ref{e:Q_b}). Points are coloured according to the initial gas fraction of the simulation. The chaos fraction clearly increases with increasing gas fraction, in contrast to
$Q_{\mathrm{b}}$ with its weak inverse dependence on increasing gas fraction.}
\label{f:Qb}

\end{figure}

Following this, we plot the chaotic fraction vs. the axisymmetric bar strength $Q_{\mathrm{b}}$ in Fig.~\ref{f:Qb}. Here, it is evident that the rising chaos cannot be attributed to an increase in this measure of bar strength as found in \cite{manos_regular_2011}.

\begin{figure}
\includegraphics{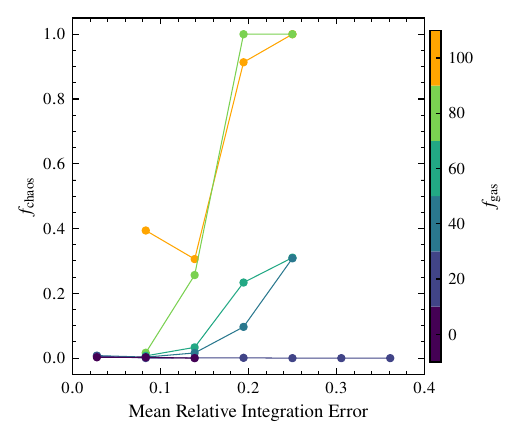}
\caption{Fraction of chaotic orbits in the complete sample vs. the mean relative integration error experienced by a star (computed as integrated minus actual N-body state normalised by orbit arc). Bins with less than 100 stars are omitted. Curves are coloured according to the initial gas fraction of the simulation.}
\label{f:error}

\end{figure}

Finally, we present the integration error diagnostic in Fig. \ref{f:error}, in which two characteristics are observed. First, for each simulation with $f_{\rm gas} > 20\%$ there is a rise in chaotic fraction with error; i.e cases where integration between subsequent snapshots mismatched the next N-body state are generally more chaotic. Second, at the same errors the higher gas fraction curves are also more chaotic. These observations persist if orbits are separated by their bar support classification.

\section{Discussion}
\label{s:dis}

\subsection{Origin of chaos}
\label{s:ori}

The physical origin of the enhanced chaos in gas-rich models warrants careful discussion. We first address potential numerical sources before turning to physical mechanisms. 
With respect to the $T_r = 83 $ Myr limit used to identify bar supporters, we track the radial periods of stars whose apocenters fall inside $R_{\rm bar}$. For simulations with $f_{\rm gas} > 0\%$, we find almost no stars whose radial period exceeds 83 Myr and have apocenters located within $R_{\rm bar}$. For the gas-free simulation that possesses the largest bar, $0.03\%$ of stars within $R_{\rm bar}$, and $41\%$ of stars in the total ($10^5$ stars) sample have less than 6 apocenters and are thus dropped to avoid contaminating either the bar or non-bar population. 
This means every bar is correctly resolved, and the total pool in the $f_{\rm gas} =0\%$ simulation is smaller with $\approx60000$ classified particles.

A known source of orbit divergence in N-body simulations is small-scale force irregularity from close particle encounters \citep{goodman_exponential_1993}. A subtlety of our implementation is that \agama\ potentials are smooth by construction, which suppresses chaos arising from such irregularity \citep[cf.][]{asano_exponential_2026}. However, because we reset the phase-space coordinates of all integrated particles at every snapshot to those of the live N-body simulation (which retains local force irregularity from the AMR grid and no inherent smoothing scale is imposed), each reset introduces an instantaneous correction to the deviation vectors.

To assess whether this procedure artificially inflates the chaotic fraction, we track the per-particle integration error and compare it against $f_{\rm chaos}$ in Fig.~\ref{f:error}. At fixed integration error, gas-rich simulations are systematically more chaotic, arguing against a purely numerical origin. The figure also offers a physical interpretation: particles whose dynamics deviate most from the smooth mean-field potential, i.e. those encountering genuine force irregularities, tend to be the most chaotic across all models. The one caveat is that if the smooth potential fits the particle distribution progressively more poorly at higher gas fractions, the measured chaos could be partly spurious; we cannot fully exclude this possibility.

Physical mechanisms have been proposed to link changes in a potential to enhanced chaos. \cite{kandrup_transient_2003} showed that rapid oscillations of a gravitational potential can induce transient chaos when they resonate with orbital frequencies, and that this can occur even if frequencies don't match perfectly. In our models, the turbulent gas disc provides a rapidly evolving potential, and we argue it is the dominant driver of the elevated chaotic fractions rather than bar properties alone. This is supported by Fig.~\ref{f:Qb}, which shows a negative correlation between $f_{\rm chaos}$ and the bar strength $Q_{\mathrm{b}}$, in contrast to the strong correlation found by \cite{manos_regular_2011} using the SALI and GALI methods applied to analytical barred-galaxy potentials. This figure is consistent with progressively weaker bars as one increases the gas fraction (cf. Fig \ref{f:Rbar}). The turbulent gas therefore appears to dominate over bar dynamics and, indeed, directly shapes the emerging bar properties \citep{bland-hawthorn_turbulent_2024}.

\begin{figure*}

\includegraphics[width=\textwidth]{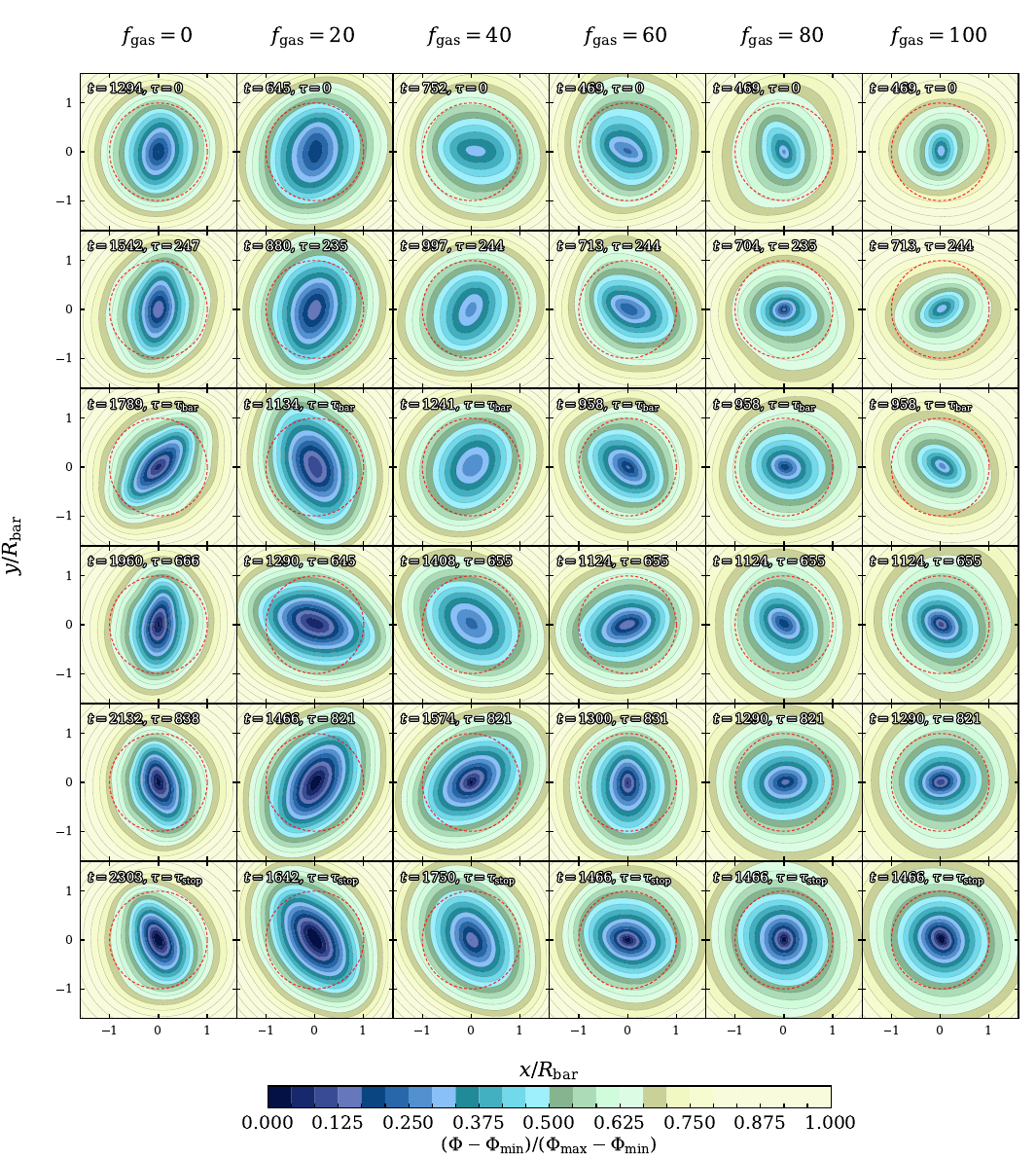}
\caption{Evolution of the total gravitational potential $\Phi$ for each model throughout 6 snapshots within the classification runs. The simulation time $t$ and the time since reference snapshot $\tau$ are annotated in Myr. For each $f_{\rm gas}$, $\Phi$ is normalised to its range over the run. Each physical scale is also normalised to $R_{\rm bar}$, which is measured at $\tau_{\rm bar}$ and indicated as a red unit circle. This implies the same bands are not expected to have similar morphology between models, for they correspond to different energies. Note how $\Phi$ deepens at the highest gas fraction as it becomes rounder; this is a consequence of violent relaxation, i.e. the mechanism driving the orbital restructuring.}
\label{f:potentials}

\end{figure*}

A further mechanism involves the build-up of a central mass concentration (CMC) - from  secular build-up through gas inflow or a supermassive black hole \citep{kormendy_secular_2004}  - this can suppress bars by populating chaotic orbits \citep{hasan_chaotic_1990, pfenniger_dissipation_1990, norman_bar_1996}. This is {\it a priori} consistent with the deepening of the central potential seen in Fig.~\ref{f:potentials}. However, monitoring the enclosed mass within multiple apertures ($R_{\rm bar}/10$, $0.05$, $0.1$, and $0.25$~kpc) relative to the reference snapshot reveals no CMC whose mass grows systematically with gas fraction. This is consistent with \cite{shen_destruction_2004}, who showed that bar destruction requires a CMC that is both highly centrally concentrated and of order a few percent of the disc mass. We note, nonetheless, that the observed rapid deepening of the potential may act dynamically as an effective CMC even in the absence of a discrete mass concentration. 

Finally, \cite{merritt_chaos_1996} showed that chaotic mixing can erode triaxiality, and that a fluctuating potential promotes this mixing only insofar as it promotes the chaos. Our results are broadly consistent with \cite{athanassoula_bar_2013} in finding that bars weaken with increasing gas fraction, though we find no evidence that the buckling instability plays a role; higher gas fractions instead render the disc more robust against such disruption \citep{bla26a}. Notably, bars in the \nexus\ models form {\it faster} with increasing $f_{\rm gas}$ (Table~\ref{t:ref}), as discussed in detail in \cite{bland-hawthorn_turbulent_2024}, suggesting that the gas simultaneously accelerates bar formation and hastens its subsequent weakening through chaotic mixing.

\subsection{Bar to bulge transformation}
\label{s:relaxation}

As we have shown, the most gas-rich bars in our simulations experience a transformation that we have studied from different perspectives. The final values of the axis ratios in Fig. \ref{f:axis} reveal that these systems evolve into a more oblate configuration by about 1 Gyr. The transition to an oblate rotator seems to follow soon after the chaotic fraction ($f_{\rm chaos}\gtrsim 50\%$) begins to dominate in these models. How are we to understand this?

To understand the rapid collapse of an emerging/dissolving bar into an oblate rotator, we appeal to a classical argument from stellar dynamics. In statistical mechanics, a gas relaxes to a Maxwell-Boltzmann distribution through two-body collisions. But stellar systems are collisionless, i.e. the two-body relaxation time $\sim N\tau_{\rm dyn}/(8\ln N)$, where $\tau_{\rm dyn}$ is the orbit's crossing time, exceeds the age of the universe by orders of magnitude \citep{jea28,cha42}. How is rapid relaxation achieved?

In a seminal paper, \citet{lyn67} argued for a `violent relaxation' process with a dynamical timescale ($\sim$crossing time) that is much shorter than the two-body timescale. During the initial collapse, the gravitational potential fluctuates violently and rapidly. One manifestation of this is the strong baryon sloshing in the same simulations on kpc scales \citep{bland-hawthorn_turbulent_2025}, a phenomenon recently confirmed by the {\tt FIRE} team \citep{myr26}. A star moving through a fluctuating potential does not conserve its energy. It can gain or lose energy depending on whether the potential deepens or shallows as it passes through \citep{zhang_enhanced_2026,mod26}. 

We refrain from reproducing Lynden-Bell's statistical mechanics since the steps are founded on the Fermi-Dirac distribution function (DF) familiar to  physics undergraduates. Crucially, his argument is based on maximising entropy in an incompressible phase space; the latter plays the role of the Pauli exclusion principle. By Liouville's theorem, the collisionless Boltzmann equation preserves phase-space density. Due to exclusion, no two stars can occupy the same phase-space element. Maximising entropy subject to this constraint drives the system towards its most probable state \citep{dej86}. Interestingly, violent relaxation exploits the phase-space density and is therefore independent of particle mass, unlike relaxation in a Maxwell-Boltzmann gas. Thus, in the absence of other influences, all stellar populations evolve to the same radial distribution. This is completely consistent with the particle-mass invariance we find in Fig. \ref{f:axis} and Fig. \ref{f:result}. 

Crucially, this relaxation is most effective when orbits are chaotic and ergodic
\citep[e.g.][]{sar25}: such an orbit fills its accessible phase-space volume completely, whereas a regular orbit is confined to an invariant torus and contributes only via phase mixing. Any surviving regular orbit families therefore contribute to incomplete relaxation. The fluctuating potential of a turbulent, gas-rich disc promotes precisely the chaotic orbital structure needed to complete the mixing, which is what we observe at the highest gas fractions (see Sec.~\ref{s:ori}).

A chaotic orbit is ergodic but its phase space is bounded by conserved quantities. In a rotating disc, the relevant conserved quantity is the azimuthal angular momentum $L_z$, which is an exact integral of motion in an axisymmetric potential. At fixed energy $E$ and fixed $L_z$, the accessible configuration space is an annulus bounded by an inner and outer radius set by the centrifugal barrier and the potential well respectively. The time-averaged density of a chaotic orbit is therefore an oblate shell, not a sphere. A spherical endpoint would require an isotropic $L_z$ distribution, but the dissolving bar and its surrounding disc carry strong net $L_z$ that chaotic mixing cannot erase. 

The endpoint is therefore fully determined. The most probable state consistent with the conserved distributions of $E$ and $L_z$ is described by a distribution function $f(E, L_z)$, which is necessarily axisymmetric and oblate. This is Lynden-Bell's prediction but for a rotating collisionless system undergoing violent relaxation, and it is almost certainly what we are observing at the highest gas fractions, i.e. an oblate rotator whose flattening reflects the angular momentum inherited from the bar population that chaotic mixing has randomised in every degree of freedom except $L_z$.

Since the degree of oblateness of the relaxed system is characterised by the ratio of ordered rotational 
to random kinetic energy, one can write down an approximate equation for the degree of flattening, i.e.
\begin{equation}
\label{e:trot}
    \frac{T_{\rm rot}}{T_{\rm rand}} = 
    \frac{\langle L_z^2 \rangle / 2\langle R^2 \rangle}
         {\langle v_R^2 + v_z^2 \rangle / 2},
\label{e:ob}
\end{equation}
for which $T_{\rm rot} = \langle L_z^2\rangle / 2\langle R^2\rangle$ is the mean (specific) rotational kinetic energy assuming a flat rotation curve; the denominator $T_{\rm rand} = \langle v_R^2 + v_z^2\rangle/2$ 
is the random (specific) kinetic energy in the meridional plane, excluding any ordered azimuthal 
streaming.\footnote{This equation is closely related to the line-of-sight $(v/\sigma)_{\rm LOS}$ aperture measurement used in the analysis of galaxy surveys to separate disc-like from spheroidal galaxies \citep[e.g.][]{wis25}.} 
$L_z = R\,v_\phi$ is the azimuthal angular momentum per unit mass, $v_\phi$ is the azimuthal velocity, $v_R$ and $v_z$ are the radial 
and vertical velocity components respectively, and angle brackets denote averages over 
the stellar population. (The factors of 2 divide out but they embody a correction for the degrees of freedom.) All of the quantities in Eq.~\ref{e:ob} are functions of an intrinsic radius. 
Note that this equation can be expressed in terms of 3D stellar dispersions $\sigma_i$ ($i=R, \phi, z$) if we substitute $\sigma_i^2 = \langle v_i^2\rangle - \langle v_i\rangle^2$. 

In Table \ref{t:T}, we present the median $T_{\rm rot}/T_{\rm rand}$ in the last $200$ Myr of each simulation computed from the full sample of $\sim 10^5$ stars within an aperture of $R < R_{\rm bar}$. 

After \citet{bin77}, we can write
\begin{equation}
T_{\rm rot}/T_{\rm rand}=\sqrt{\varepsilon/(1-\varepsilon)}=\sqrt{(1-c/a)/(c/a)}
\end{equation}
which is zero when $c/a=1$, an unflattened non-rotating sphere. 
In Table \ref{t:T}, $T_{\rm rot}/T_{\rm rand}$ is near unity for $f_{\rm gas} = 80\%, 100\%$ (shown in bold), and this is when $c/a\approx 0.5$, as we see in Fig. 3 by the end of the simulation. The high $f_{\rm gas}$ ($\gtrsim  80\%$) simulations are consistent with an oblate rotator. 

For $f_{\rm gas} < 80\%$, $c/a<0.5$ and $T_{\rm rot}/T_{\rm rand}>1$ as observed. The low $f_{\rm gas}$ simulations are more dominated by rotation due  to the bar. In Fig. 3, $c/a=0.25$ gives $T_{\rm rot}/T_{\rm rand}\approx 1.7$; these stars have more rotation and did not experience violent relaxation where a lot of energy is lost. The $T_{\rm rot}/T_{\rm rand}$ values are as expected for $f_{\rm gas} = 20\%, 60\%$ but seem small for $f_{\rm gas} = 0\%, 40\%.$

\begin{table}
    \centering
    \caption{The median ratio of  rotational to random energy (Eq. \ref{e:trot}) late in the simulations from all sampled stars within $R_{\rm bar}$.}
    \label{t:T}
    \begin{tabular}{cc}
        \hline
        $f_{\rm gas}$ & $T_{\rm rot}/T_{\rm rand}$\\
        \hline
        0   & 1.10\\
        20  & 1.78\\
        40  & 1.17\\
        60  & 1.87\\
        {\bf 80} & {\bf 0.97}\\
        {\bf 100} & {\bf 0.97}\\
        \hline
    \end{tabular}
\end{table}

\section{Conclusions}
\label{s:conclusions}

Using the \nexus\ N-body/hydrodynamic framework, we have investigated the orbital structure of stellar bars forming in idealised disc galaxies across the full range of initial gas fractions ($f_{\rm gas} = 0$--$100\%$), motivated by the prevalence of barred systems observed at $z \gtrsim 4$ with JWST and ALMA. The central question is whether the turbulent, gas-rich environments characteristic of high-redshift discs fundamentally alter the dynamical character of their bars compared to their low-redshift counterparts.

Bar-supporting stars are identified through a libration test in the rotating bar frame, and their orbits are classified as regular or chaotic using the Smaller Alignment Index (SALI) method applied to live N-body trajectories, a novel implementation that tracks secular evolution rather than extrapolating frozen snapshots. The key result is a sharp transition near the initial gas fraction of $f_{\rm gas} \approx 60\%$: below this, chaotic fractions remain close to zero throughout the bar's lifetime; above it, the chaotic fraction rises rapidly to $60$--$85\%$ within roughly 1~Gyr. This transition coincides with, and precedes, the collapse of the bar into an oblate, rotationally supported bulge, a transformation absent in less gas-rich models.

The origin of the enhanced chaos is attributed to the rapidly evolving gravitational potential driven by the turbulent gas disc rather than to bar strength itself, since no correlation between $f_{\rm chaos}$ and the bar strength parameter $Q_{\mathrm{b}}$ is found. No significant central mass concentration is detected, though the observed deepening of the central potential may act dynamically in this fashion.

The bar collapse is interpreted through the \cite{lyn67} violent relaxation framework. As chaotic orbits ergodically fill their accessible phase space, the system evolves toward the maximum-entropy distribution function $f(E, L_z)$ consistent with its conserved energy and angular momentum. This interpretation is strengthened by the observed mass independence of the bar axis ratios and chaotic fraction. Because the dissolving bar carries strong net $L_z$ that chaotic mixing cannot erase, the relaxed endpoint is necessarily an oblate rotator rather than a sphere.

This process may represent a universal pathway for rapid bulge formation in the gas-rich, turbulent conditions of the early Universe. The bar-driven route demonstrated here operates on a timescale of order 400–800 Myr.  To our knowledge, this is the shortest formation timescale reported for a central stellar bulge in an initially stable disc \citep[cf.][]{dek14}. By comparison, the clump-migration mechanism of \citet{nog99,nog00}, in which massive disc fragments spiral inwards through dynamical friction, requires at least several Gyr.

The statistical physics governing why these two routes produce such different timescales is discussed by \citet{bla26a}.

The epoch at which stellar bulges and bars first appeared remains an open question, but the recent JWST discovery of a barred spiral with a prominent bulge at $z \approx 5.1$ \citep{wang_massive_2026} - when the Universe was only 1.2 Gyr old - places a hard lower bound on how early these structures can form and demands a closer examination of rapid formation mechanisms. Our results suggest that baryon-dominated, gas-rich discs can meet this constraint without invoking major mergers or long-term secular evolution. These issues are explored further in companion papers.

\section*{Acknowledgements}

We are grateful to Eugene Vasiliev and Romain Teyssier for continued assistance and adaptions with \agama\ and \ramses\ respectively.
TTG acknowledges financial support from the Australian Research Council (ARC) through Australian Laureate Fellowships awarded to JBH (FL140100278) and TRB (FL220100117), for partial funding from Lund University, and from the James Arthur Pollock memorial fund awarded to the School of Physics, University of Sydney. JBH also acknowledges support from the ARC grant DP220103384 shared with Ken Freeman. OA acknowledges support from the Knut and Alice Wallenberg Foundation, the Swedish Research Council (grant 2025-04892), the Swedish National Space Agency (SNSA grants 2023-00164 and 2025-00405), the LMK foundation, and eSSENCE, a Swedish strategic research programme in e-Science. C.F.~acknowledges funding provided by the Australian Research Council (Discovery Projects DP230102280 and DP250101526), and the Australia-Germany Joint Research Cooperation Scheme (UA-DAAD). JMVA acknowledges funding from The University of
Sydney International Scholarship

The computations and data storage were enabled by two facilities: (i) the National Computing Infrastructure (NCI) Adapter Scheme, provided by NCI Australia, an NCRIS capability supported by the Australian Government; and (ii) LUNARC, the Centre for Scientific and Technical Computing at Lund University (resource allocations LU 2023/2-39 and LU 2023/12-6). We further acknowledge high-performance computing resources provided by the Leibniz Rechenzentrum and the Gauss Centre for Supercomputing (grants~pr32lo, pr48pi, pn76ga and GCS Large-scale project~10391), the Australian National Computational Infrastructure (grant~ek9) and the Pawsey Supercomputing Centre (project~pawsey0810) in the framework of the National Computational Merit Allocation Scheme and the ANU Merit Allocation Scheme.
\section*{Data Availability}

The data underlying this article will be shared on reasonable request to the corresponding author.

\bibliography{sample701}{}
\bibliographystyle{mnras}

\bsp	
\label{lastpage}
\end{document}